# Testing Robustness of Camera Fingerprint (PRNU) Detectors

Fernando Martín-Rodríguez.

fmartin@tsc.uvigo.es.

atlanTTic research center for Telecommunication Technologies, University of Vigo,

Campus Lagoas Marcosende S/N, 36310 Vigo, Spain.

*Abstract-* In the field of forensic imaging, it is important to be able to extract a "camera fingerprint" from one or a small set of images known to have been taken by the same camera. Ideally, that fingerprint would be used to identify an individual source camera. Camera fingerprint is based on certain kind of random noise present in all image sensors that is due to manufacturing imperfections and thus unique and impossible to avoid. PRNU (Photo-Response Non-Uniformity) has become the most widely used method for SCI (Source Camera Identification). In this paper, we design a set of "attacks" to a PRNU based SCI system and we measure the success of each method. We understand an attack method as any processing that alters minimally image quality and that is designed to fool PRNU detectors (or, generalizing, any camera fingerprint detector). The PRNU based SCI system was taken from an outstanding reference that is publicly available.

**Keywords:** image forensics, camera identification, fingerprint, forgery, PRNU.

## I. Introduction

Camera fingerprint is based in the unavoidable imperfections present on electronic imaging sensors due to their manufacturing. These imperfections derive on a multiplicative noise that can be modeled in this equation [3], [4], [5]:

$$Im_{out} = (I_{ones} + Noise_{cam}).Im_{in} + Noise_{add} \quad (1)$$

Where $Im_{in}$ is the "true" image presented to the camera (the incident light intensity), $I_{ones}$ is a matrix of ones, $Noise_{cam}$ is the "sensor noise", "**.**" is matrix point by point product and $Noise_{add}$ is additive noise from other sources.

Most SCI methods are based on somehow estimating $Noise_{cam}$ and comparing results from different images.

An attacking method would be any processing over $Im_{out}$ that "to some extent" accomplishes the following two purposes:

- Not affecting the visual appearance of $Im_{out}$ to a great extent.
- Making it more difficult to estimate $Noise_{cam}$ from the new processed image.

Similarly to the importance of cryptanalysis to cybersecurity world, study on camera fingerprint attacks is important to image forensics because it is the only way to demonstrate robustness of proposed extraction and comparison methods.

The remainder of this paper is organized as follows: first, we describe briefly the PRNU fingerprint and its computation; then we describe the attack methods used; next, we describe the tests implementation followed by their results.

## II. PRNU Camera Fingerprint

PRNU stands for Photo-Response Non-Uniformity and it is used to refer to sensor noise as modeled by equation 1. PRNU estimation has become the "de-facto" standard for camera fingerprint. In [14] an extensive list of methods is compared and PRNU is acknowledged as the most used method and the one most present on literature. PRNU estimation is always based on some kind of image denoising filter to estimate $Im_{in}$ in equation 1. Having a set of images (known to be captured by the same camera), a PRNU pattern can be estimated and then compared to any image from an unknown camera. Comparison must yield a high positive value if patterns are similar, meaning that it is probable that source camera is the same. Nowadays, despite of recent advances in this field, it is very difficult to come to a legally irrefutable result.

The denoising process used to compute PRNU can be of very different types, for example:

- Simply a 3x3 median filter like in [6].
- The well-known Wiener filter [7].
- Modifications of Wiener filter, for example in [1] (implementation chosen as basis [2]), they use an implementation on the WVT (Wavelet Transform [8]). This means using the Wiener formula ($H = \frac{Image}{Image+Nois}$) in the WVT transform, where we assume white noise (constant in the WVT). Multiplying image WVT by 1-H and erasing wavelet lowpass component, we have the WVT of PRNU pattern (image).

Last but not least, it is important to use a good comparison method to detect similarities between the different extracted patterns. Following [1,2], we will use PCE (Peak to Correlation Energy,). PCE has demonstrated good performance in PRNU comparison [9]. PCE is the maximum of the cross correlation between the extracted PRNU noise for a given image and that same image multiplied by the PRNU pattern of the class (camera) to be verified.

## III. ATTACKING METHODS

Let's first, establish a classification on methods and then, we will comment the different methods that have been included in the test section. Method categories could be summarized as follows:

- <u>Noise addition (or modification)</u>: ramdomizing least significant components (bits) does not add or reduce noise but it merely modifies it. Nevertheless, it is the first idea that comes up, fingerprints are based on noise and modifying noise might work.
- <u>Geometric distortion</u>: geometric distortions like pixel position scrambling and/or rotating and de-rotating image (with a slight angle error) have had success against other "noise like" patterns like watermarks.
- <u>Noise reduction</u>: If we reduce image noise, we will also erase the fingerprint (at least to some extent).
- <u>Combined methods</u>: methods constructed cascading two of more of the previous ones (may be from the same category or not).

### A. Noise addition (or modification)

We have implemented two flavors of this idea. First, modifying the **n** least significant bits in the image (pixel) domain; second, adding noise on the "Discrete Cosine Transform" (DCT) coefficients. In the second case, Watson matrix from JPEG standard [10] is used to determine the allowed noise quantity. Noise is a uniform random variable with different interval on each DCT position.

### B. Geometric distortions

We have implemented three methods from this idea:

- Scrambling pixels: moving them to a nearby, random position (we define a maximum radius, **r**, to maintain it controlled). Gaussian distribution is used to scramble pixels.
- Rotating and de-rotating: image is rotated a significant angle (say $\alpha$=15º). Pixel "bicubic" interpolation is forced. Then image is "de-rotated" (rotated again an angle of $-\alpha+\beta$, where $\beta$ is a small error, say 0.50º). Bicubic interpolation is forced again.
- Scaling and de-scaling: image is up-scaled by a significant factor (say **sf**=3). "Lanczos3" interpolation [13] is forced. Then we erase first line and first column. Image is downscaled to its original size, forcing a "non-uniform sampling" and using "Lanczos2" interpolation.

### C. Noise reduction

We have implemented two methods for this. First, it is the standard Wiener filter. Second, it is a Wiener implementation in the wavelet transform domain inspired in the PRNU extraction method [1].

### D. Combined methods

As it is easily deduced, we call combined methods to those constructed cascading two or more of the previously described method. We have implemented the following combinations (they were inspired by our first tests on individual methods trying to combine the strengths of each):

- Combination of simple noise addition and geometric techniques. No Wiener or other noise reduction (**n**=3, **r**=2, $\alpha$=10º, $\beta$=0.5º, **sf**=3).
- Wiener filtering first, rotation and de-rotation ($\alpha$=10º, $\beta$=0.5º), followed by a "deblurring" method for improving image quality. We used Lucy-Richardson deconvolution filter [11].

## IV. DESIGN OF TESTS

Tests were performed using a two photo collections. First we used a self-created collection of 80 photos from four different cameras. Afterwards, we repeated the tests with images from the "Dresden Image Database" [15], we used 36 images from 6 different cameras. All photos were JPEG files generated by each device with NO processing. As we encountered different image resolutions, we preprocessed the images simply cropping a centered square sub-image of size 2048x2048 (thus, condition for adding a camera into the studio is that resolution of the minimum image dimension is at least 2048 pixels). In the first database, we included images from smartphones. In some cameras (concretely this is typical of smartphones), images taken with a rotated sensor (vertical format: height greater than width) are automatically rotated to be in vertical format. In this case, we rotate them 90º so that images are always in horizontal format. Besides of testing PRNU algorithms robustness, these tests also detect which kind of cameras are better (or worse) for fingerprint detection.

Attacking algorithms were implemented in MATLAB [12]. Tools were designed to perform tests that work on the condition that images from each camera are saved in a separate directory. Therefore, adding a new camera to the test is very easy.

For each camera the first $n_t$ images ($n_t$ = ¼ of total images for that camera) are used to create a PRNU pattern using tools in [2]. After that:

- We compute a confusion matrix for all cameras (understood as the truth table resulting from trying to identify the source camera for all images that are NOT used in computing the PRNU). This allows us measuring the "camera identification error rate" before any attack is applied.
- We apply each of the attacks defined in section 3 and we re-compute the confusion matrix and error rate.
- We also compute the SNR (Signal to Noise Ratio) after the attack (considering original and attacked, or noisy, images) as a means for considering image quality degradation. We present the average SNR for each method or "attack". We also check the "visual" quality of processed images.

With this experiment, we generate the tables in section 5, which are used to draw conclusions on PRNU performance and to assess best attacking methods.

## V. TEST RESULTS

For the test we used two databases. The first one is made of images from four cameras: Canon EOS 1100D (low cost DSLR), Sony A5000 (mirrorless camera), Canon PowerShot SX710 (bridge camera) and Huawei P20lite (smartphone). The second one is the well known "Dresden Image Database" [15], we selected images from six cameras: Canon Ixus 70 (2 cameras of this model), Casio Ex150 (also two instances) and Kodak M1063 (two instances).

With no attack, camera identification error rate is 21% for the first database and 25% for the second one. All the attacks make this rate go higher. Note that with four cameras, a "random identification" algorithm should yield a 25% correct identification rate (a 75% error rate). This means that an attack able to produce an error rate of 75% or higher is a successful attack. With 6 cameras, the successful attack must come to 83% (100-100/6).

We have measured error rate in two ways for each method. First, we have done the training (computation of PRNU) with original (non-attacked) images and, second, we have trained with attacked images. Notice that forensics people can have the suspicious camera or not. Sometimes [5], PRNU is used for clustering: trying to find out if some images come from the same camera (with no access to it). Error rate is always computed with images different from training ones.

We summarize the tests results in table 1 (next page). See that we have included two columns about image quality. This is important because an attack is not "useful" if it is very "noticeable" (if image quality is severely damaged). We have assessed final image quality in two ways: subjective and objective (SNR computation). See that for some attacks (for example rotating or scrambling), subjective quality can be high but SNR gets low because pixels are moved and SNR is computed comparing pixels at the same position. Subjective quality has been checked with opinions of seven volunteers.

See that, generally speaking, attack success is higher when training is done with "non-attacked" images. Perhaps this is because when PRNU is computed after "being altered by an attack", it will contain these alterations and it will be more difficult to be fooled. For example, with the rotating method, the error rate goes down from 67% to 43% (first database) when training with attacked (fooled) images. Probably, detected noise pattern is rotated by the attack and PRNU robustness is much higher if it can train with "rotated" noise. We also acknowledge that PRNU is more resilient that we have thought. See that only the last combined method is able to reach (in both databases) the "success" ratios of 75%/83% and only when training with non-attacked images.

We have also studied briefly the fingerprint robustness according to the different kinds of camera. The error rate is always less in the smartphone images whereas maximum error appears in the mirrorless camera (that it is also the newest and more expensive one). This seems to confirm the "a priori" idea that fingerprint is easier to be extracted in less quality (more noisy) sensors.

In table 1, we include a column with mean execution time for each attack. See that MATLAB implementation is not efficient, besides a fast attack implementation is not the main purpose of this work but the possibility and effectiveness of attacks. Tests were run in an Intel I7-6700HQ (2.60 GHz), 8.00 Gb RAM.

In figure 1, we can see a graphic example. We have an original image from the Canon Powershot camera (first database). This image is processed (attacked) with the last procedure of table 1: Wiener filtering cascaded with rotation and a de-blurring filter. Then we compute PRNU patterns from both images. Notice than in ordinary tests, more than one image is used to get the PRNU pattern and then this pattern is correlated to those extracted from input images for trying to guess the source camera. In this case, we only want to notice the change between extracted patterns. This kind of patterns is not very visual to be shown because they are merely amplified noise. They consist on real valued matrixes of the same size of original image with values from -16 to 16. In this case, we have represented a truncated version of matrix absolute value. Interpretation is simply that brighter points are pixels where we have been able to extract a significant value for PRNU. See that for the original image, algorithm fails to compute pattern only in some white (perhaps over-exposed) pixels. For the processed image, computing PRNU pattern becomes much more difficult.

We have compiled more examples of processed images in the next URL:
https://www.flickr.com/photos/189133275@N08/collections/72157715180732621/

## VI. THE "DEFINITE" METHOD

After the test results, if we were prompted to select one, that one would be the last one of table 1: Wiener filtering + rotation/de-rotation + de-blurring filtering. That method was used to construct example in figure 1. As we can notice in that figure the only drawback in this method is the presence of some "artifacts" on image borders (see top-left corner of the attacked image). This is due to the (0.5º) error in de-rotation. There are some pixels that "are not defined" and that have been computed with a linear extrapolation filter yielding this result.

We have designed a new (simple) algorithm that avoids this problem. The method is simply:
- Cropping image to erase "non-defined" border pixels.
- Re-escaling image to original size.

Results for this new method have been recomputed yielding:
- First database: 17 dB SNR (but better visual quality), 19 seconds execution time, 77% error rate for non-fooled training, 48% error rate for fooled images training.
- Second database: 16 dB SNR (but better visual quality), 19 seconds execution time, 83% error rate for non-fooled training, 54% error for fooled images training.

See figure 3 for more details.

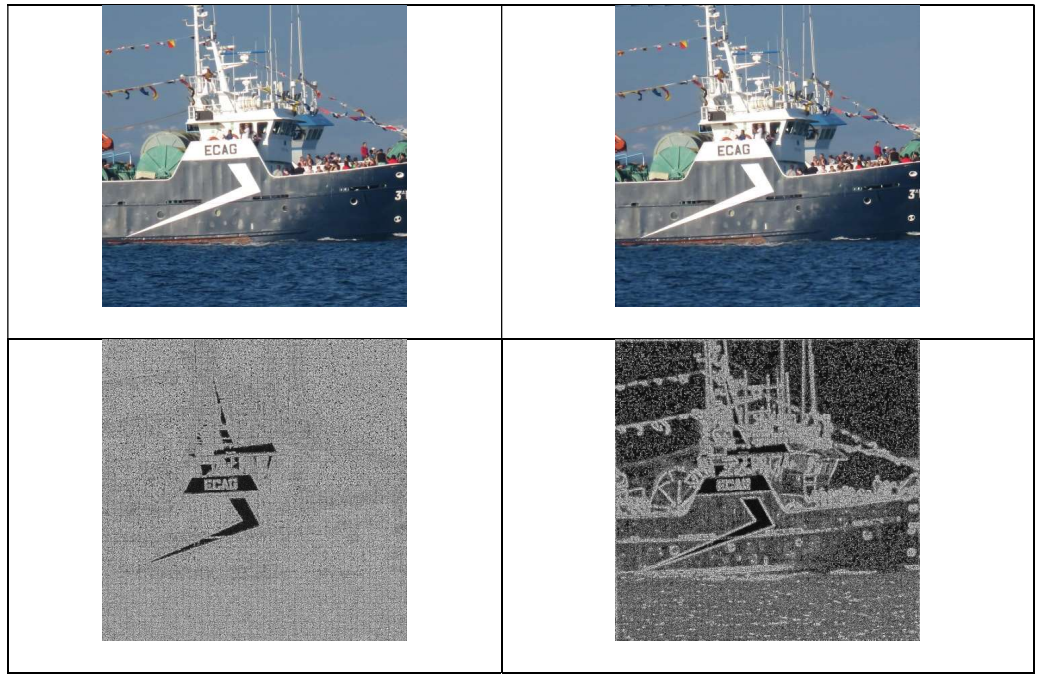

Fig. 1. <u>Above</u>: original image on the left, image processed by the wiener+rotation method on the right. <u>Below</u>: noise patterns (PRNU) obtained from each of the above images. PRNU computation is not good in dark points (images below) which correspond to overexposed pixels in original image and are more present in processed image. Normally PRNU patterns are computed (trained) from several original images.

TABLE I
Summary of tests results (X/Y ➔ Self-created database/Dresden database)

| ATACKING METHOD: | Visual Quality: | Mean SNR quality (dB): | Mean execution time for attack (s): | <<Error Rate>>, Non fooled training set (%). | <<Error Rate>>, Fooled training set (%). |
|---|---|---|---|---|---|
| Aleatorizing least significant bits (**n**=3). | Good, except color degradations (clouds). | 38 | 11 | 25/33 | 31/38 |
| Introducing noise on DCT coefficients. | Good | 49 | 22 | 22/29 | 25/29 |
| Scramble randomly pixels (r=1). | Good | 30 | 15 | 28/25 | 32/58 |
| Rotating and de-rotating (A=10º,a=0.5º). | Good, except artifacts on borders. | 19 | 11 | 67/83 | 43/42 |
| Scaling and de-scaling (sf=3). | Good | 42 | 12 | 27/29 | 25/38 |
| Ordinary wiener filter. | Good | 32 | 2 | 43/54 | 51/58 |
| Wavelet transform wiener filtered and inverted. | Good | 41 | 3 | 30/33 | 37/45 |
| Combination of simple noise addition and geometric techniques (n=3, r=2, A=10º, a=0.5º, sf=3). | Good-, artifacts in some borders, quantification in color degradation areas (sky, clouds). | 20 | 16 | 73/75 | 70/58 |
| Wiener + dotating/derotating + deblurring (Lucy) | Good | 19 | 19 | 83/88 | 68/58 |

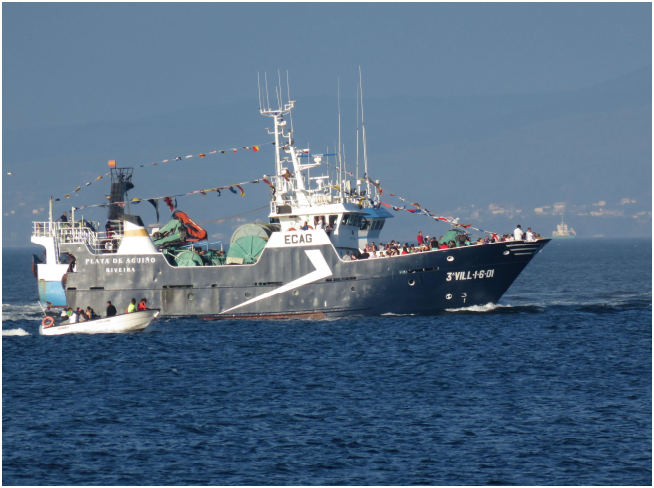

Fig. 2. Confusion matrix samples. <u>Left</u>: no attack. <u>Right</u>: last combined method.

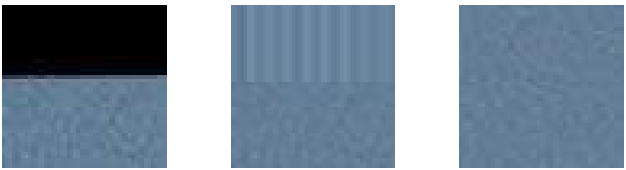

Fig. 3. Above: image of figure 1 processed with the new (definite) method, below: border details with "no correction method", "linear filtering" and new method.

## VII. Conclusions and future lines

We have explored the robustness of the nowadays most used method for camera fingerprint identification (PRNU) and we have concluded that it has strong resilience against simple attacks like those based on geometric operations and/or based on altering noise. The application of more elaborated methods based on noise theory like Wiener filtering allows creating much better attacks. Best attack in our study is created combining the two best single methods tested previously.

Future lines of research would be developing new methods and studying more deeply their interactions with fingerprint computation. This would also allow developing more robust fingerprint methods.